\documentclass[10pt,conference]{IEEEtran}

\makeatletter
\newcommand*{\rom}[1]{\expandafter\@slowromancap\romannumeral #1@}
\makeatother
\usepackage{amsfonts,epsfig,graphicx,tabularx,multirow,fancyhdr,amsmath,subfigure}
\usepackage{amsmath,epsfig,graphicx,setspace,float}
\usepackage{pstricks,colortab,pifont}
\usepackage{lscape,multicol,longtable,subfigure,soul}
\usepackage[noadjust]{cite}
\usepackage[small,bf]{caption}
\usepackage{gensymb}
\usepackage{subfigure}
\usepackage{algorithmic}
\usepackage{algorithm}

\usepackage{graphicx,color}
\usepackage{amsmath,epsfig,cite,subfigure, amssymb}
\usepackage{color}
\usepackage{subfig}
\usepackage{times}
\usepackage{mathptm}
\usepackage{booktabs}
\usepackage{float}
\usepackage{setspace}
\begin{document}

\title{Utility based Joint Uplink Downlink Optimization Energy Harvesting enabled Hybrid RF/VLC system}

\author{\IEEEauthorblockN{Shivanshu Shrivastava}}
\maketitle

\begin{abstract}
Hybrid Radio Frequency/ Visible Light Communication (RF/VLC) systems have been recognized as effective means to enhance data rate. So far the study of data rate has been limited to downlink communications. In this letter, we study the maximization of the total data rate which also involves the uplink data rate. Simulation results verify the convergence of the proposed algorithm.
\end{abstract}

\section{Introduction}
The increasing demand for high data rate after the mobile internet revolution has brought the need for alternatives to radio frequency (RF) communications. Visible Light Communication (VLC) has emerged as an efficient candidate in this regard. It uses illumination from the deployed light sources for data transmission, offering several advantages like high data rate, lesser interference with the co-existing RF devices, efficient frequency reuse, efficient unregulated spectrum usage and power saving by accomplishing dual objectives of illumination and data transmission. It uses the light sources in the room as access points (APs). These APs are provided data through a backhaul network and they transmit information to the mobile terminals (MTs) present in the indoor environment. 


The primary objective in VLC is maximization of the achievable data rate to achieve throughput maximization. In \cite{paper_four}, the authors have studied a typical hybrid RF/VLC network with user association   based on the minimum distance criteria and equal bandwidth allocation among UEs. In \cite{paper_one}, the authors have studied joint load balancing and optimal power allocation for achievable rate maximization, but without considering optimality in bandwidth allocation. 
In \cite{paper_three}, a dual hop system with energy harvesting  has been studied to obtain the optimal fraction of time a UE should be connected to the AP to minimize the packet loss probability. 
\cite{paper_six} studies the joint user association and power allocation issue where the optimal association is obtained by the UE channel conditions  while considering fairness among UEs. 
\cite{assoc_1} studied  the user association problem with lighting constraints for a VLC only system and proposed a greedy algorithm maximizing the SINR based utility function. 
 In \cite{assoc_2}, a novel association scheme (termed ``anticipatory association scheme") was proposed which tries to anticipate the future locations of the UE and adjust association according to that. 
 
 The primary limitation in the above works is that only downlink systems have been studied. The total performance of an MT depends on its uplink transmission also. However, to the best of our knowledge, the achievable rate point of view of the uplink transmission has not been studied till now. The objective of this paper is to perform the joint optimization of the downlink-uplink of the MTs to maximize the total achievable rate of the VLC system.
 
 The primary requirement of maximizing the uplink throughput is the transmission power of the MT.
Recently light energy harvesting has emerged as an efficient mechanism for the MTs for their uplink transmission. In \cite{energy_harvested} the authors perform the outage probability analysis for light energy harvesting. In \cite{secure},  the secrecy outage probability is studied for an light energy harvesting system. The secrecy outage probability analysis is done for VLC systems facing attack from the eavesdroppers. The eavesdropper steals data from the transmitting AP intended for a legitimate MT. The authors do the analysis for the secrecy outage probability. The limitation of these works is that the energy harvesting has not been studied in the light of achievable rate maximization.

In this paper, we perform the study the joint uplink-downlink achievable rate maximization of a VLC system. This significant research topic has remained unstudied and the present work is the first study which considers the joint optimization of the achievable rate in the uplink and downlink communication in VLC systems. The joint uplink-downlink study leads to a study from the utility-gain point of view of an MT. It is evident that the utility point of view demands maximization of the achievable rate in the uplink and the downlink. We consider that the MTs are enabled with time splitting based energy harvesting. The harvested energy is used by the MTs for their uplink data transmission. The contributions of our work are enlisted as 

The contributions of our work have been enlisted as follows:
\begin{itemize}
\item We design a novel uplink downlink joint optimization problem for the hybrid RF/VLC set up. The MTs harvest energy during the downlink transmission from the VLC signals in the time splitting mode and use this energy for data transmission in the uplink. 
\item We show that the joint uplink downlink problem is convex in the time splitting coefficient for fixed configuration of bandwidth and power. 
\item We propose a simplified Lagrange based algorithm to solve the above optimization problem.
\item We obtain the optimal time splitting coefficient for energy harvesting which maximizes the total achievable sum rate. 
\end{itemize}
The rest of the paper is organized as follows:
In Section \rom{2}, we elaborate on the system model. In Section \rom{3}, we formulate the joint optimization problem for the VLC system. In Section \rom{4}, we propose a simple solution for the formulated problem. In Section \rom{5}, we present the simulation results and Section \rom{6} concludes the paper.

\section{System Model}\label{SysMod}
Consider $i=1,2,\ldots,K$ light sources deployed on the ceiling of a typical room as shown in Fig.\ref{model}. Each light source acts as a VLC AP. Let there be $j=1,2,\ldots,M$ MTs present in the room. 
We consider the MTs at a height $h$ from the floor. 
Each VLC AP has its illumination region. The $j$th MT lying in the illumination region of the $k$th VLC is said to be within its field of view (FOV). We assume that if LOS components between the k$th$ VLC AP are unblocked for the $j$th MT, it receives an SINR more than the minimum threshold from the $k$th VLC AP. 
If the SINR falls below the threshold value, it  switches its connection to another VLC AP which gives an SINR more than the threshold. 
The downlink communication to an MT is done through the VLC APs while the uplink communication from the MT is done to the RF AP with RF signals.  
We consider the MTs equipped with colocated energy harvesting (EH) and information decoding (ID) receivers. 
We assume a backhaul circuit connected between the RF and the VLC APs for underlying operations and data transmission between them. The downlink data is transmitted from the VLC APs to the MTs. We consider a non-coordinated transmission. After receiving the data from the APs, the MTs transmit their data to the APs on the uplink. 
We now enlist the notation of all the parameters with their physical meanings in this set up for clarity in the evaluations ahead:

\begin{itemize}
\item Downlink datarate from a VLC AP to an MT: $R_d$
\item Downlink communication time: $T_d$
\item Uplink datarate from an MT to the RF AP: $R_u$
\item Energy harvested by a typical MT from the incoming downlinkn signals: $E_H$
\item Uplink communication time: $T_u$
\item Transmission power of an MT for uplink transmission: $P_H=E_H/T_u$
\item Time splitting coefficient: $\alpha$. 
Note that during the downlink communication, information decoding will be done for $\alpha$ fraction of $T_d$ while energy harvesting will be done for $(1-\alpha)$ fraction of $T_d$.
\item $B_v$ VLC channel BW
\item $B_r$ RF channel BW
\item $P_t$ transmit power of the VLC AP assumed to be same for all the APs. 
\end{itemize}
We next describe the downlink and the uplink channel gains. 

\begin{figure}[h]
\centering
\includegraphics[height=7cm,width=9cm]{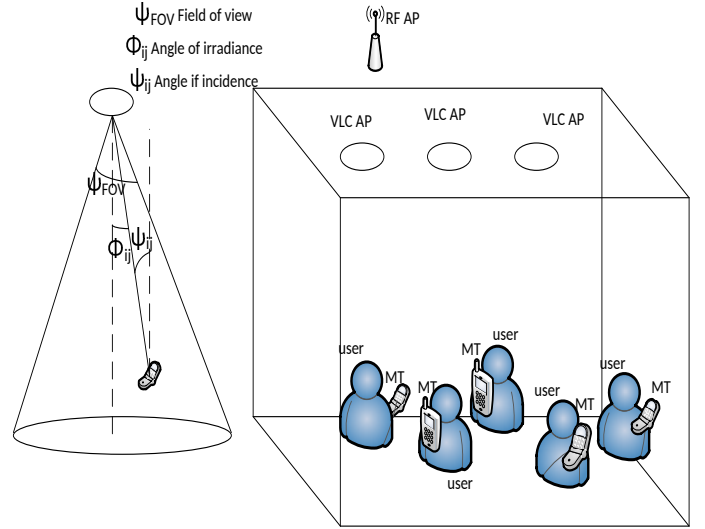}
\caption{Hybrid RF/VLC system}
\label{model}
\end{figure}

\subsection{The Downlink Communication:} The VLC APs transmit the data to the MTs on the donwlink. As mentioned before, the VLC APs transmit through dimming of the visible LED light which is unnoticeable to the eyes. The channel gain for this communication is defined by the Lambertian model. The downlink channel gain from the $i$th VLC AP to $j$th MT is given as:
\begin{equation}
{G_{ij}} = \frac{{(m_{ij} + 1)A\rho {{\cos }^{m_{ij}}}{\phi _{ij}}\cos {\psi _{ij}}{T_s}({\psi _{ij}})g({\psi _{ij}})}}{{2\pi d_{ij}^2}}
\end{equation}
where ${{T_s}(\psi _{ij} )}$ is the gain of the optical receiver filter and is  unity or a constant value within the field of view (FOV) of a receiver, $m_{ij}$ is the Lambertian emission factor given by $m_{ij} =  - \frac{1}{{{{\log }_2}(\cos {\phi _{1/2}})}}$, $\phi _{ij}$ is the angle of irradiance, and $\psi _{ij}$ is the angle of incidence, $d_{ij}$ is the distance between $i$th AP and the $j$th MT and
\begin{equation}
g(\varphi ) = \frac{{{n^2}}}{{{{\sin }^2}{\varphi _{FOV}}}}
\end{equation}
\begin{equation}
n = \frac{\text{{speed of light in vaccum}}}{{\text{speed of lightin that optical material}}}
\end{equation}

The Downlink rate is formulated as follows:
\begin{equation}\label{dlRate}
{R_d} =  {B_v}{\log _2}\left( {1 + \frac{{{P_T}{G_{ij}}}}{{{N_0} + \sum\limits_{k \ne i} {{G_{kj}}} }}} \right)
\end{equation}
The MTs are equipped with energy harvesting capability. The MTs harvest energy from the incoming downlink signals. We neglect the harvest from the ambient light in the present analysis. The
Energy harvested by a typical MT is formulated as
\begin{equation}
{E_H} = {C_{jRF}}{T_d}{\rho _j}\left( {\frac{{{{\left( {{P_T}} \right)}^2}}}{{d_{ij}^4}}{{\cos }^{2{m_i}}}({\phi _j})(1 - \alpha ) + \sum\limits_{k \ne i} {\frac{{{{\left( {{P_T}} \right)}^2}}}{{d_{kj}^4}}} {{\cos }^{2{m_k}}}({\phi _j})} \right)
\end{equation}
where $C_{jRF}$ is the coefficient of optical power to RF power conversion, $\rho _j$ is the optical to electrical conversion efficiency coefficient.\\

\subsection{ Uplink:} The uplink channel gain for the short range communication, particularly in the indoor scenarios, is highly likely to have line of sight (LOS) components. These channels are modeled with Rician distribution as done in \cite{secure, energy_harvested}. On the uplink channel, the signal envelop from $j$th MT to RF AP is considered as Ricean faded with the probability density function (pdf):
\begin{equation}
f(r) = \frac{{2r(1 + K)}}{\Omega }\exp \left( { - K - \frac{{{r^2}(1 + K)}}{\Omega }} \right).{I_0}\left( {2r\sqrt {\frac{{K(1 + K)}}{\Omega }} } \right)
\end{equation}
where $\Omega$ is the variance of the signal, $K$ is Rician factor which
corresponds to the ratio of the power of the LOS (specular)
component to the average power of the scattered component,
and $I_0 (·)$ is the modified Bessel function of the first kind and
0th order.
 


Uplink rate:
\begin{equation}
{R_u} = {B_v}{\log _2}\left( {1 + \gamma ({E_H})} \right)
\end{equation}

where $\gamma ({E_H})$ is the uplink SNR.
\begin{equation}
\gamma ({E_H}) = \frac{{{P_H}|{h_{jm}}{|^2}}}{{{N_0}d_j^n}} = \frac{{{E_H}|{h_{jm}}{|^2}}}{{{T_u}{N_0}d_j^n}} = \frac{{\left( {(1 - \alpha ){K_1} + {K_2}} \right)|{h_{jm}}{|^2}}}{{{T_u}{N_0}d_j^n}}
\end{equation}
where ${K_1} = {C_{jRF}}{T_d}{\rho _j}\frac{{{{\left( {{P_T}} \right)}^2}}}{{d_{ij}^4}}{\cos ^{2{m_i}}}({\phi _j})$ and ${K_2} = {C_{jRF}}{T_d}{\rho _j}\sum\limits_{k \ne i} {\frac{{{{\left( {{P_T}} \right)}^2}}}{{d_{kj}^4}}} {\cos ^{2{m_k}}}({\phi _j})$.
Note that in RF uplink SNR, loop interference (LI), co-channel interference come into picture. Loop interference happens on the node which is receiving and transmitting simultaneously. It is the interference of the received signal on the transmitted signal at the node. In RF uplink communication literature, LI is considered other than co-channel interference. However [1] and [2], the authors have not considered it. Co channel interference is also not considered. The SNR expression is only noise limited. As of now we start with the model in [1] and [2].

\section{Problem Statement}
It can be seen that higher the magnitude of $\alpha$, lower is $E_H$, $P_H$ and thus $R_u$, while  higher the value of $R_d$. Thus $\alpha$ creates a trade-off between $R_d$ and $R_u$. 

Secondly, we assume the MTs to be energy starved. Hence a minimum value of energy has to be harvested to perform uplink transmission. Mathematically,

\begin{equation}\label{prob_1}
\begin{array}{l}
 \mathop {\max }\limits_\alpha  \quad R = \alpha {R_d} + {R_u} \\ 
 st\quad \quad 0 \le \alpha  \le 1 \\ 
 \end{array}
\end{equation}
Elaborating the problem as:
\begin{equation}\label{prob_2}
\begin{array}{l}
 \mathop {\max }\limits_\alpha  \quad \alpha {B_v}{\log _2}\left( {1 + \frac{{{P_T}{G_{ij}}}}{{{N_0} + \sum\limits_{k \ne i} {{G_{kj}}} }}} \right) +  \\ 
 \quad \quad {B_v}{\log _2}\left( {1 + \frac{{\left( {(1 - \alpha ){K_1} + {K_2}} \right)|{h_{jm}}{|^2}}}{{{T_u}{N_0}d_j^n}}} \right) \\ 
 \end{array}
\end{equation}

\section{Proof of convexity in $\alpha$}
In this section, we prove the convexity of the formulated utility function in $\alpha$ as follows:
\begin{equation}
\begin{array}{l}
 R = \alpha {B_v}{\log _2}\left( {1 + \frac{{{P_i}{G_{ij}}}}{{{N_0} + \sum\limits_{k \ne i} {{P_k}{G_{kj}}} }}} \right) +  \\ 
 {B_R}{\log _2}\left( {1 + \frac{{\left( {(1 - \alpha ){K_1} + {K_2}} \right)|{h_{jm}}{|^2}}}{{{T_u}{N_0}d_j^n}}} \right) \\ 
 \end{array}
\end{equation}
We use some dummy variables to make the calculations easier as $x = \alpha,   a = {{\rm{P}}_i}{{\rm{G}}_{ij}}, {b_1} = {B_v}, {b_2} = {B_R}, b = {N_0}{B_v}, c = \sum\limits_{k = 1}^K {{P_T}{G_{kj}}}, d = {K_1}|{h_{jm}}{|^2}, e = {K_2}|{h_{jm}}{|^2}$, and $g = {T_u}{N_0}d_j^n $. The problem can be written as

\begin{equation}\label{rate}
R = \alpha {b_1}{\log _2}\left( {1 + \frac{a}{{b + c}}} \right) + {b_2}{\log _2}\left( {1 + \frac{{\left( {(1 - \alpha )d + e} \right)}}{g}} \right)
\end{equation}

\begin{equation}\label{first_der}
\frac{{dR}}{{d\alpha }} = {b_1}\log \left( {\frac{a}{{c + b}}} \right) - \frac{{{b_2}d}}{{g\left( {\frac{{d(1 - \alpha ) + e}}{g} + 1} \right)}}
\end{equation}
The second derivative of $R$ is obtained as
\begin{equation}\label{first_der}
\frac{{{d^2}R}}{{d{x^2}}} =  - \frac{{{b_2}{d^2}}}{{{{(dx - g - d - e)}^2}}}
\end{equation}

The second derivative of (\ref{rate}) shows that $R$ is concave in $\alpha$ and for any fixed bandwidth and power, there indeed should exist an $\alpha$ which maximizes $R$. 

%
\section{Optimal $\alpha$}
In this Section, we solve for the optimal $\alpha$ for a fixed bandwidth and transmission power. It can be seen that the problem in (\ref{prob_1}) is subject to a constraint that $\alpha$ must lie between 0 and 1, as it is the time fraction of the energy harvesting to the transmission time. For fixed resources, this problem is a scalar function problem. We need to optimize only one parameter and the constraint is also on the same parameter.  The problem in (\ref{prob_1}) can be written as 
\begin{equation}\label{prob_2}
\begin{array}{l}
 {\cal{P}}:\mathop {\max }\limits_\alpha  \quad \alpha {B_v}{\log _2}\left( {1 + \frac{{{P_T}{G_{ij}}}}{{{N_0} + \sum\limits_{k \ne i} {{G_{kj}}} }}} \right) + {B_v}{\log _2}\left( {1 + \frac{{\left( {(1 - \alpha ){K_1} + {K_2}} \right)|{h_{jm}}{|^2}}}{{{T_u}{N_0}d_j^n}}} \right) \\ 
 st\quad \quad 0 \le \alpha  \le 1 \\ 
 \end{array}
\end{equation}
The problem ${\cal{P}}$ can be written as 
\begin{equation}\label{prob_ll}
\begin{array}{l}
 {\cal{P}} : \mathop {\max }\limits_\alpha  \quad \alpha {B_v}{\log _2}\left( {1 + \frac{{{P_T}{G_{ij}}}}{{{N_0} + \sum\limits_{k \ne i} {{G_{kj}}} }}} \right) +  \\ 
 {B_v}{\log _2}\left( {1 + \frac{{\left( {(1 - \alpha ){K_1} + {K_2}} \right)|{h_{jm}}{|^2}}}{{{T_u}{N_0}d_j^n}}} \right) \\ 
 st\quad \quad g_1(\alpha)= \alpha - 1  \le 0 \\ 
 \quad \quad  g_2(\alpha)=-\alpha  \le \;0 \\ 
 \end{array}
\end{equation}
The Lagrange of problem ${\cal{P}}$ can be written as
\begin{equation}\label{prob_L}
{\cal{L}}(\alpha ,\lambda ,\mu ) = R + \lambda (\alpha  - 1) + \mu \alpha
\end{equation}
The KKT conditions of the problem in (\ref{prob_L}) can be expressed as
\begin{equation}\label{kkt_1}
\frac{{dR(\alpha )}}{{d\alpha }} + \lambda  + \mu  = 0
\end{equation}

\begin{equation}\label{kkt_2}
\lambda  \ge 0,\mu  \ge 0
\end{equation}

\begin{equation}\label{kkt_3}
 - \alpha  \le 0
\end{equation}

\begin{equation}\label{kkt_4}
\alpha  - 1 \le 0
\end{equation}

\begin{equation}\label{kkt_5}
\mu \alpha  = 0
\end{equation}

\begin{equation}\label{kkt_6}
\lambda (\alpha  - 1) = 0
\end{equation}
From simple algebraic substitutions, $\alpha$ can be written as
\begin{equation}
\alpha  = 1 - \left( {\frac{{{b_2}}}{{{b_1}\log \left( {\frac{a}{{c + b}}} \right) + \lambda  - \mu }} - \frac{{e - g}}{d}} \right)
\end{equation}
Combining the conditions given in expressions (\ref{kkt_1}) - (\ref{kkt_6}), the solution for $\alpha$ is given as 
\begin{equation}\label{eq_final_1}
\left\{ \begin{array}{l}
 \alpha  = \max \left( {0,1 - \left( {\frac{{{b_2}}}{{{b_1}\log \left( {\frac{a}{{c + b}}} \right) + \lambda  - \mu }} - \frac{{e - g}}{d}} \right)} \right) \\ 
 \,\;\; \le 1 \\ 
 \end{array} \right.
\end{equation}
The above expression can be efficiently solved with the bisection method on $\lambda$ and $\mu$ as shown in the following algorithm.

\begin{algorithm}[ht!]
\label{algo_1}
\caption{Optimal value of $\alpha$}
\begin{algorithmic}
\STATE{ $l=0; \epsilon > 0$; select a feasible $\alpha^{(0)}$}
\WHILE{$||{\alpha ^{(l + 1)}} - {\alpha ^{(l)}}|| > \varepsilon$ }
\STATE{$l=l+1$}
\STATE{Compute $\lambda$ and $\mu$}
\STATE{Compute $\alpha$ according to (\ref{eq_final_1})}
\ENDWHILE
\end{algorithmic}
\end{algorithm}


\section{Simulation Results}
In this section, we plot the results showing the variation of the total utility of the UE under the different settings of the resources. It can be seen that the utility of the UE maximizes at the optimal value of $\alpha$ obtained in (\ref{alpha1}).
\subsection{Convexity of the Total Sum Rate}
It can be clearly seen in (\ref{prob_1}) that the first term of the sum rate expression is the increasing function of $\alpha$. On the other hand, $R_u$ is the decreasing function of $\alpha$. This is quite intuitive from the fact that higher the value of $\alpha$, lower is the $E_H$ and hence lower is available transmission power for uplink. Thus for fixed bandwidth and transmission power, $R$ should be concave in $\alpha$ which is shown in Fig. ()


\begin{figure}
\begin{tabular}{cc}
\small \scriptsize\bf(a)&\scriptsize\bf (b)\\
\includegraphics[width=0.25\textwidth,trim=5 0 0 22,clip]{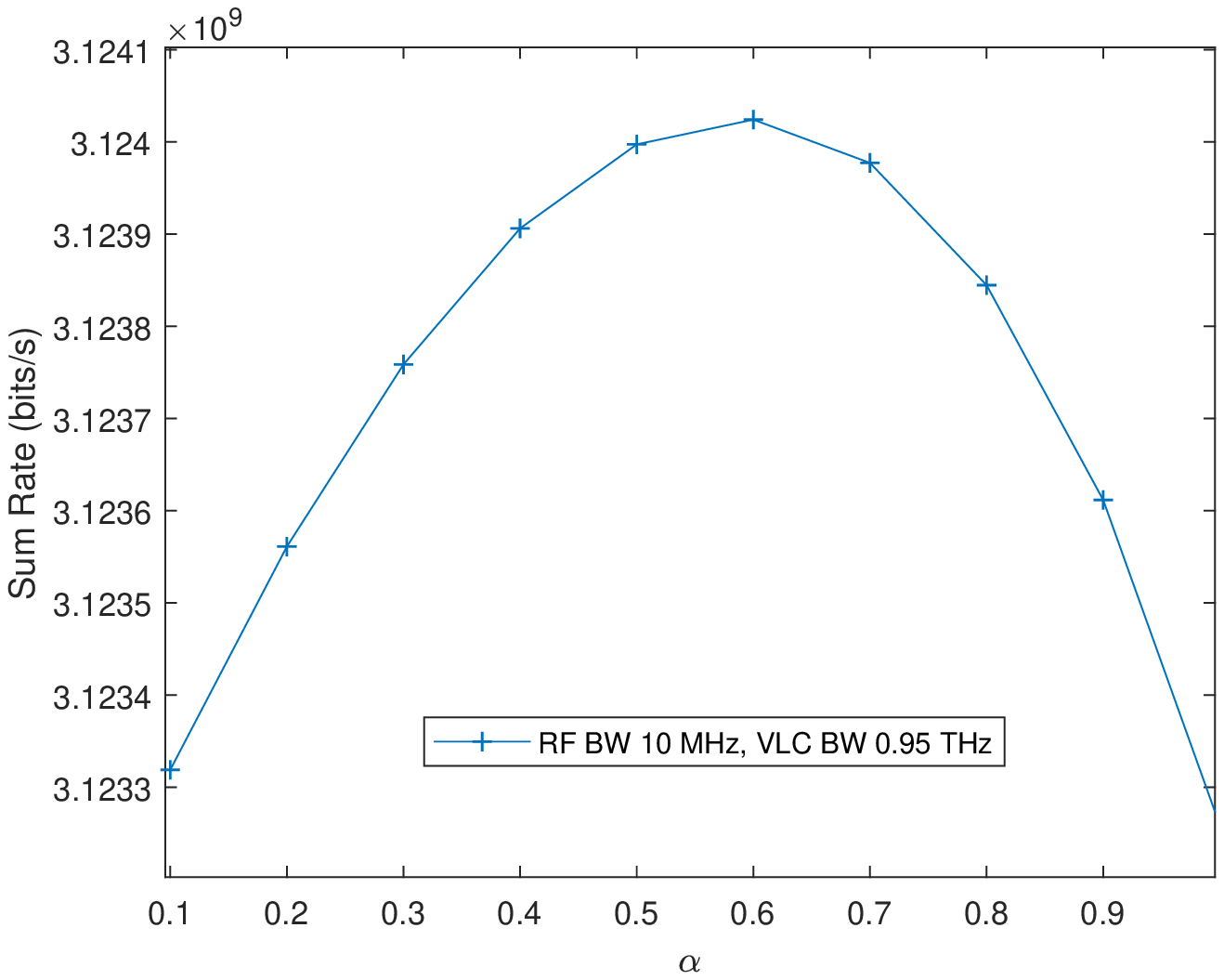}&
\includegraphics[width=0.2\textwidth]{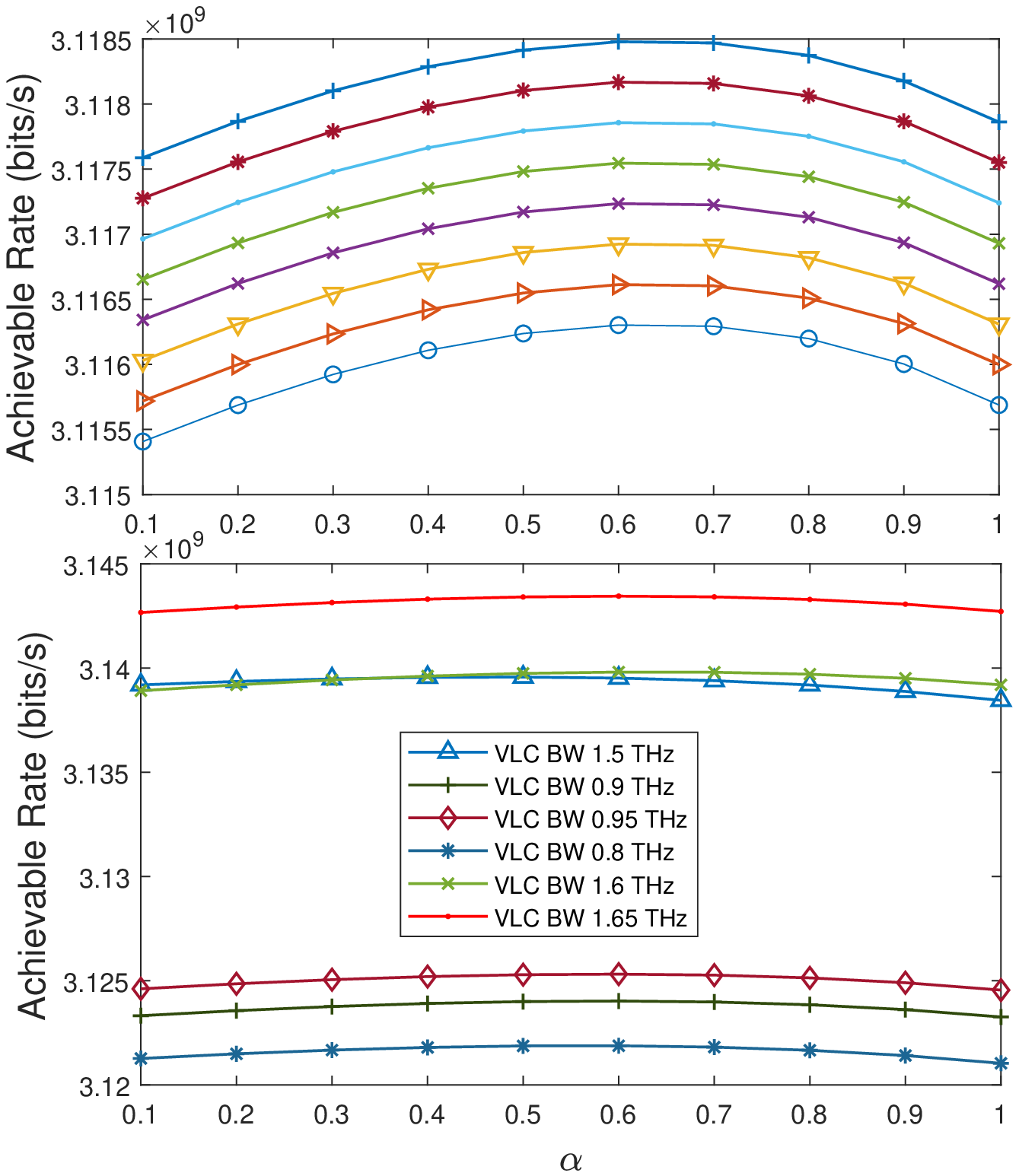}
\end{tabular}
\caption{Convexity of Achievable Rate with $\alpha$ (a) For one Bandwidth Configuration (b) For Different RF and VLC Bandwidth configurations }
\label{fig:tenuserD}
\end{figure}

\begin{figure}[h]
\centering
\includegraphics[height=7cm,width=9cm]{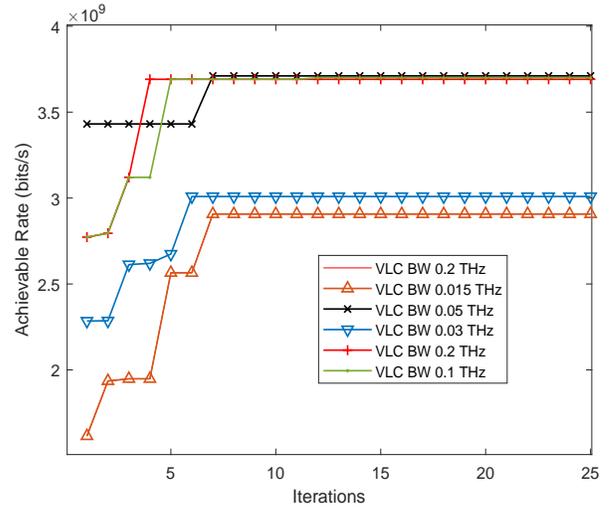}
\caption{Figure depicting the convergence of the algorithm for different VLC bandwidths}
\label{model}
\end{figure}

\section{Conclusion}
This letter reports a novel study on the joint uplink-donwlink achievable rate maximization of a VLC communication. We obtained the optimal decoding time coefficient to maximize the sum rate in the uplink-downlink. With the help of numerical simulations, we prove that the optimal value of the time decoding coefficient maximizes the sum rate in the uplink-downlink.

\end{document}